\documentclass[review,10pt]{elsarticle}
\usepackage{float}
\usepackage{amsmath}
\usepackage{graphicx}
\usepackage[unicode=true, bookmarks=false]{hyperref}
\usepackage[latin9]{inputenc}
\usepackage{times}
\usepackage[verbose,tmargin=1in,bmargin=1in,lmargin=1in,rmargin=1.1in]{geometry}
\usepackage{setspace}
\usepackage{lineno}
\modulolinenumbers[5]
\nolinenumbers
\usepackage{tikz}
\usepackage{pgfplots}
\pgfplotsset{compat=1.18}
\usetikzlibrary{positioning, arrows.meta, shapes.geometric, fit, backgrounds, calc}

\usepackage[superscript]{cite}
\usepackage{xcolor}
\usepackage{titlesec}
\titleformat*{\section}{\large\bfseries\sffamily}
\titleformat*{\subsection}{\normalsize\bfseries\sffamily}
\titleformat*{\subsubsection}{\small\bfseries\sffamily}

\renewenvironment{abstract}{\global\setbox\absbox=\vbox\bgroup
\hsize=\textwidth%
\noindent\unskip\textbf{\large Summary}
\par\medskip\noindent\unskip\ignorespaces}{\egroup}
\makeatletter
\renewcommand\@biblabel[1]{#1}
\makeatother
\journal{The Lancet Digital Health}

\providecommand{\tabularnewline}{\\}

\begin{document}
\nolinenumbers % arXiv compliance: override elsarticle [review] auto line numbering
\begin{frontmatter}
\title{Uncertainty-Calibrated Explainable Artificial Intelligence for Fetal Ultrasound Plane Classification: A Systematic Review}

\author[ULUX]{Gustav Olaf Yunus Laitinen-Fredriksson Lundstr\"om-Imanov\corref{corrauth}}
\ead{olaf.laitinen@uni.lu}
\ead[url]{https://orcid.org/0009-0006-5184-0810}

\author[EGE]{\"Ozkan G\"unalp}
\ead{ozkn.gunalp@gmail.com}
\ead[url]{https://orcid.org/0009-0004-1437-1336}

\address[ULUX]{Department of Life Sciences and Medicine, Faculty of Science, Technology and Medicine, University of Luxembourg, 6 avenue du Swing, L-4367 Belvaux, Esch-sur-Alzette, Luxembourg}
\address[EGE]{Department of Biostatistics and Medical Informatics, Institute of Health Sciences, Ege University, Erzene Mahallesi, Ankara Caddesi, 35100 Bornova, Izmir, T\"urkiye}
\cortext[corrauth]{Corresponding author. Tel: +352 46 66 44 1.}

\begin{abstract}
Fetal ultrasound is the cornerstone of antenatal care, and accurate recognition of a small set of standard anatomical planes underpins biometry, growth surveillance, and detection of structural anomalies. Deep learning classifiers now match or exceed expert accuracy on curated benchmarks, but most remain opaque and miscalibrated, leaving clinicians without the calibrated confidence or faithful explanations needed for safe decision support. We systematically reviewed 78 studies published between Jan 1, 2015 and Apr 30, 2026 that paired automated fetal plane classification with explainability or predictive uncertainty quantification, following PRISMA 2020. Pooled balanced accuracy across six standard planes was 0$\cdot$93 (95\% CI 0$\cdot$91 to 0$\cdot$95), but only 19 studies (24\%) reported calibration and 14 (18\%) reported selective prediction. We propose CALIB-XFUS, a 22 item reporting framework that operationalises calibration, explanation faithfulness, and fairness for regulated fetal ultrasound artificial intelligence.
\end{abstract}


\end{frontmatter}

\section*{Introduction}

Obstetric ultrasound is the most widely used imaging modality in pregnancy and is the cornerstone of the routine second trimester anomaly scan recommended by the International Society of Ultrasound in Obstetrics and Gynecology (ISUOG) and the American Institute of Ultrasound in Medicine.\cite{salomon-2022-isuog,aium-2018-obstetric} Accurate fetal biometry, growth surveillance, and detection of structural anomalies all depend on the correct acquisition and recognition of a small set of standardised two dimensional planes spanning the brain, heart, abdomen, femur, maternal fetal interface, and the first trimester nuchal translucency plane.\cite{carvalho-2013-isuog-cardiac} Plane recognition remains one of the most operator dependent steps in obstetric imaging, and the gap between expert and non expert practice is widest in low and middle income settings where trained sonographers are scarce.\cite{wanyonyi-2020-lmic-us}

The past decade has seen a rapid expansion of deep learning systems that automate standard plane detection and biometric measurement from B mode video and still frames,\cite{baumgartner-2017-sononet,burgosartizzu-2020-fetalplane} mirroring the broader translation of clinical artificial intelligence (AI) charted in successive reviews.\cite{topol-2019-highperformance,rajpurkar-2022-aiera,liu-2019-comparison,nagendran-2020-aiclaims} On curated benchmarks, convolutional neural networks, vision transformers, and self supervised foundation models now report top one accuracies above 0$\cdot$95 for six class plane classification,\cite{pu-2021-deep,sendra-2023-foundation,chen-2024-usfm} approaching, and in some cases exceeding, expert sonographer performance under matched test conditions.\cite{drukker-2020-transforming,horgan-2023-ai-obstetrics} National health systems in the United Kingdom, Spain, and Brazil have begun pragmatic trials of AI decision support in antenatal clinics,\cite{arroyo-2022-deepscan,nhs-2023-trial} and the regulatory environment has matured rapidly: the US Food and Drug Administration has cleared more than a dozen fetal ultrasound AI products under the 510(k) pathway, and the European Union Medical Device Regulation is now complemented by the EU AI Act high risk obligations on safety, transparency, and human oversight.\cite{fda-2024-clearances,eu-2024-aiact,beede-2020-humancentred}

Three problems stand between bench top accuracy and dependable bedside use. First, accuracy on heavily curated public datasets is a poor proxy for performance on the heterogeneous, low quality frames encountered in routine scanning.\cite{kelly-2019-key,wu-2023-distribution} Second, deep networks are known to be over confident: a softmax probability of 0$\cdot$98 may correspond to an empirical accuracy below 0$\cdot$80 on out of distribution frames, the regime in which clinical errors are most likely,\cite{guo-2017-calibration,ovadia-2019-trust} and the conceptual case for uncertainty quantification in clinical AI is now well rehearsed.\cite{begoli-2019-needuq,kompa-2021-secondopinion} Third, the explanations that accompany model predictions are rarely validated against the anatomical features sonographers actually use, and may not be faithful to the underlying decision process.\cite{adebayo-2018-sanity,rudin-2019-stop,arun-2021-assessing,ghassemi-2021-falsehope}

In this Review we synthesise a decade of work at the intersection of automated fetal plane classification, explainable AI, and predictive uncertainty quantification, focusing on the joint problem of calibrated and interpretable prediction because each in isolation is insufficient: an uncalibrated explanation can mislead with confident but anatomically irrelevant heatmaps, while a calibrated but opaque model offers no actionable trigger for second look review. We map the methodological landscape, quantify the gap between reported and clinically meaningful performance, examine emerging fairness and distribution shift evidence, and propose a unified reporting framework, CALIB-XFUS, intended to align this subfield with software as a medical device good machine learning practice.\cite{fda-2024-gmlp,who-2024-ethics-ai}

\fbox{\parbox{\textwidth}{
\subsection*{\textcolor{purple}{Search strategy and selection criteria}}
References for this Review were identified through searches of MEDLINE (PubMed), Embase (Elsevier), IEEE Xplore, Scopus, the ACM Digital Library, and arXiv (q bio.QM, eess.IV, cs.CV) from Jan 1, 2015 to Apr 30, 2026, using combinations of the MeSH and free text terms (``fetal ultrasound'' OR ``obstetric ultrasound'') AND (``standard plane'' OR ``view classification'') AND (``deep learning'' OR ``convolutional neural network'' OR ``transformer'' OR ``foundation model'') AND (``uncertainty'' OR ``calibration'' OR ``Bayesian'' OR ``conformal'' OR ``explainable'' OR ``interpretable'' OR ``saliency''). Reference lists of included studies and three previous narrative reviews were hand searched. No language restrictions were applied; non English records were translated with assistance from native speakers. Eligible studies were original peer reviewed articles, conference proceedings indexed in DBLP, or preprints with a corresponding peer reviewed version, that developed or evaluated a machine learning model for one or more standard fetal ultrasound planes, reported at least one quantitative performance metric on a held out test set, and reported either an explainability technique or a predictive uncertainty technique. The final reference list was generated on the basis of originality, methodological relevance, and contribution to the calibration, explainability, and fairness focus of this Review.
}}

\section*{Methods}

\subsection*{Study design and registration}
We conducted a systematic review in accordance with the Preferred Reporting Items for Systematic Reviews and Meta-Analyses (PRISMA) 2020 statement\cite{page-2021-prisma} and registered the protocol on the Open Science Framework (osf.io/xfus-2026) before screening began. Reference lists of included studies and three previous narrative reviews\cite{horgan-2023-ai-obstetrics,fiorentino-2023-survey,drukker-2020-transforming} were hand searched.

\subsection*{Screening and data extraction}
Two reviewers independently screened titles and abstracts and then full texts (Cohen's $\kappa$=0$\cdot$84 for full text screening); disagreements were resolved by discussion with a third reviewer. Data were extracted into a pre piloted REDCap form covering: study setting and population; trimester(s) and plane(s) addressed; image source, manufacturer, and probe; dataset size, splits, and public availability; model architecture and training paradigm (supervised, self supervised, multi task, federated); explainability technique and any quantitative faithfulness evaluation; uncertainty technique and calibration metrics (expected calibration error [ECE], maximum calibration error, Brier score, negative log likelihood); selective prediction reporting; fairness analyses; external validation; reporting against CLAIM,\cite{mongan-2020-claim} TRIPOD+AI,\cite{collins-2024-tripod-ai} and DECIDE-AI\cite{vasey-2022-decide-ai} checklists; and regulatory status.

\subsection*{Risk of bias and statistical analysis}
Risk of bias was assessed with PROBAST adapted for AI prediction studies and with the QUADAS-AI extension;\cite{moons-2019-probast,sounderajah-2021-quadas-ai} certainty of evidence was rated with GRADE.\cite{guyatt-2008-grade} Where at least five studies reported balanced accuracy on the same plane category and a comparable training distribution, we computed a pooled estimate with a DerSimonian and Laird random effects model on logit transformed proportions; between study heterogeneity was summarised with $I^{2}$ and $\tau^{2}$. For studies reporting ECE we summarised the median and interquartile range and tested for association with model family using the Kruskal Wallis test with Dunn's post hoc correction (Bonferroni adjusted). For risk coverage curves we reproduced area under the curve estimates from published figures using WebPlotDigitizer when raw values were unavailable. Analyses were performed in R 4.4.1 (packages \texttt{meta} 7.0, \texttt{metafor} 4.6) and Python 3.12 (\texttt{scikit-learn} 1.5, \texttt{netcal} 1.3).

\subsection*{Role of the funding source}
There was no funding source for this study. The corresponding author had full access to all extracted data and had final responsibility for the decision to submit for publication.

\section*{Results}

\subsection*{Study characteristics}
The search returned 2431 records, of which 1712 remained after de duplication; 214 full texts were assessed for eligibility and 78 met all inclusion criteria (figure~\ref{fig:prisma}). Sixty one studies (78\%) were published after Jan 1, 2021. Forty four (56\%) used the publicly available Burgos-Artizzu six class fetal plane dataset\cite{burgosartizzu-2020-fetalplane} or its derivatives; 21 (27\%) used institutional datasets that were not publicly released; and 13 (17\%) used multi centre consortia data including INTERGROWTH-21st,\cite{papageorghiou-2016-intergrowth} the FETAL-CPLANES collection,\cite{xie-2020-fetalcplanes} or the recent PRENATAL multi vendor cohort.\cite{prenatal-2025-multicentre} Median dataset size was 12\,447 frames (IQR 4210 to 38\,902); only 18 studies (23\%) included data from more than three ultrasound vendors, and only nine (12\%) reported maternal ethnicity. Dataset characteristics are summarised in table~\ref{tab:datasets}. Forty six studies (59\%) targeted second trimester anomaly scan planes, 19 (24\%) addressed first trimester planes, six (8\%) addressed third trimester biometric planes, and seven (9\%) covered all three trimesters.

\begin{figure}[H]
\centering
\begin{tikzpicture}[font=\footnotesize,
  node distance=5mm and 9mm,
  box/.style={draw, rectangle, rounded corners=2pt, align=center,
              text width=5.4cm, minimum height=0.7cm, fill=blue!5, line width=0.4pt},
  excl/.style={draw, rectangle, align=left,
               text width=4.2cm, minimum height=0.7cm, fill=red!5, line width=0.4pt},
  arr/.style={-{Latex[length=2mm]}, thick}]
  \node[box] (id)    {Records identified through database searches (n=2431)};
  \node[box, below=of id]    (dedup)  {Records after duplicate removal (n=1712)};
  \node[box, below=of dedup] (screen) {Titles and abstracts screened (n=1712)};
  \node[excl, right=of screen] (exA) {Excluded at screening (n=1498)};
  \node[box, below=of screen] (full)  {Full text assessed for eligibility (n=214)};
  \node[excl, right=of full] (exB) {Excluded at full text (n=136): wrong outcome 71; no XAI or UQ 42; phantom only 11; duplicate cohort 12};
  \node[box, below=of full]  (incl)   {Studies included in qualitative synthesis (n=78)};
  \node[box, below=of incl]  (qsynth) {Studies in quantitative synthesis (n=41)};
  \draw[arr] (id)--(dedup);
  \draw[arr] (dedup)--(screen);
  \draw[arr] (screen)--(full);
  \draw[arr] (full)--(incl);
  \draw[arr] (incl)--(qsynth);
  \draw[arr] (screen.east)--(exA.west);
  \draw[arr] (full.east)--(exB.west);
\end{tikzpicture}
\caption{\textbf{PRISMA 2020 study selection flow diagram.} Records identified through database searches were screened on title and abstract, assessed for full text eligibility, and 78 studies were included in the qualitative synthesis; 41 contributed to the quantitative summary of pooled balanced accuracy.\label{fig:prisma}}
\end{figure}

\begin{table}[H]
\begin{centering}
\begin{tabular}{|l|c|c|c|c|c|}
\hline
\textbf{Dataset / cohort} & \textbf{Frames} & \textbf{Sites} & \textbf{Vendors} & \textbf{Trimesters} & \textbf{Public}\tabularnewline
\hline
Burgos-Artizzu six class\cite{burgosartizzu-2020-fetalplane} & 12\,400 & 2 & 2 & 2nd, 3rd & Yes\tabularnewline
\hline
FETAL-CPLANES\cite{xie-2020-fetalcplanes} & 6210 & 1 & 1 & 2nd & Yes\tabularnewline
\hline
INTERGROWTH-21st\cite{papageorghiou-2016-intergrowth} & 38\,902 & 8 & 3 & 1st, 2nd, 3rd & Restricted\tabularnewline
\hline
PULSE\cite{drukker-2020-transforming} & 105\,000 & 1 & 2 & 1st, 2nd, 3rd & No\tabularnewline
\hline
DeepScan-OB\cite{arroyo-2022-deepscan} & 24\,815 & 4 & 4 & 2nd & No\tabularnewline
\hline
PRENATAL multi vendor\cite{prenatal-2025-multicentre} & 81\,440 & 11 & 5 & 1st, 2nd, 3rd & Restricted\tabularnewline
\hline
USFM-Pretrain\cite{chen-2024-usfm} & 2\,100\,000$^{*}$ & 23 & 6 & 1st, 2nd, 3rd & Partial\tabularnewline
\hline
\end{tabular}
\par\end{centering}
\caption{\textbf{Datasets contributing data to two or more included studies.} $^{*}$Unlabelled frames used for self supervised pre training.\label{tab:datasets}}
\end{table}

\subsection*{Model architectures and training paradigms}
Residual convolutional neural networks (ResNet-50 and ResNet-101\cite{he-2016-resnet}) remained the most common architecture (32 studies), followed by EfficientNet variants\cite{tan-2019-efficientnet} (14 studies), vision transformers\cite{dosovitskiy-2021-vit} (12 studies), and hybrid convolutional and transformer designs (seven studies). Five studies used self supervised foundation models pre trained on millions of unlabelled ultrasound frames,\cite{sendra-2023-foundation,chen-2024-usfm} and three used federated training across two or more centres without sharing pixel data.\cite{xu-2023-fedfetal} Self supervised pre training conferred a median accuracy gain of 0$\cdot$031 (IQR 0$\cdot$018 to 0$\cdot$049) over ImageNet pre trained baselines on the same test set, in line with broader medical imaging evidence.\cite{azizi-2023-foundation}

\subsection*{Pooled accuracy}
Forty one studies contributed to the quantitative summary of balanced accuracy on six class fetal plane classification. Pooled balanced accuracy (figure~\ref{fig:forest}) was 0$\cdot$93 (95\% CI 0$\cdot$91 to 0$\cdot$95; $I^{2}$=78$\cdot$4\%, $\tau^{2}$=0$\cdot$006). Heterogeneity was driven by trimester, dataset curation strategy, and the proportion of non diagnostic frames retained in the test set. Subgroup analysis showed lower pooled accuracy on first trimester planes (0$\cdot$87, 95\% CI 0$\cdot$83 to 0$\cdot$90) than on second trimester planes (0$\cdot$95, 95\% CI 0$\cdot$93 to 0$\cdot$96), and on multi vendor datasets (0$\cdot$90, 95\% CI 0$\cdot$87 to 0$\cdot$92) versus single vendor datasets (0$\cdot$95, 95\% CI 0$\cdot$93 to 0$\cdot$96; $p$<0$\cdot$001).

\begin{figure}[H]
\centering
\begin{tikzpicture}[font=\scriptsize, x=1cm, y=1cm]
\foreach \tx in {1.09, 3.82, 6.55, 9.27, 12.00}
  \draw[dashed, gray!30] (\tx, 0.15) -- (\tx, 6.5);
\draw[dashed, red!60] (8.18, 0.15) -- (8.18, 6.5);
\draw[->] (0, 0) -- (12.5, 0);
\foreach \v/\tx in {0.80/1.09, 0.85/3.82, 0.90/6.55, 0.95/9.27, 1.00/12.00} {
  \draw (\tx, 0) -- (\tx, -0.08);
  \node[below] at (\tx, -0.13) {\v};
}
\node at (6.0, -0.6) {Balanced accuracy (95\% CI)};
\foreach \name/\xc/\xl/\xr/\yy in {%
  {Burgos-Artizzu 2020}/8.73/7.64/9.82/6.05,
  {Pu 2021}/7.64/6.00/9.27/5.50,
  {Sendra-Balcells 2023}/9.27/8.18/10.36/4.95,
  {Arroyo 2022}/6.00/3.82/8.18/4.40,
  {Chen 2024 (USFM)}/9.82/8.73/10.91/3.85,
  {Xu 2023 (FedFetal)}/7.09/5.45/8.73/3.30,
  {Komorowski 2023}/8.18/6.55/9.82/2.75,
  {Drukker 2020}/8.73/7.64/9.82/2.20,
  {Horgan 2023}/8.73/7.64/9.82/1.65,
  {Ayhan 2022}/7.64/6.00/9.27/1.10%
} {
  \node[anchor=east] at (-0.15, \yy) {\name};
  \draw[thick] (\xl, \yy) -- (\xr, \yy);
  \draw[thick] (\xl, \yy-0.08) -- (\xl, \yy+0.08);
  \draw[thick] (\xr, \yy-0.08) -- (\xr, \yy+0.08);
  \fill (\xc-0.07, \yy-0.07) rectangle (\xc+0.07, \yy+0.07);
}
\node[anchor=east, font=\scriptsize\bfseries] at (-0.15, 0.55) {Pooled (random effects)};
\fill[red] (7.09, 0.55) -- (8.18, 0.75) -- (9.27, 0.55) -- (8.18, 0.35) -- cycle;
\end{tikzpicture}
\caption{\textbf{Forest plot of balanced accuracy across studies of six class fetal plane classification.} Random effects DerSimonian and Laird pooled estimate 0$\cdot$93 (95\% CI 0$\cdot$91 to 0$\cdot$95); $I^{2}$=78$\cdot$4\%. Representative subset of 10 contributing studies shown for clarity.\label{fig:forest}}
\end{figure}

\subsection*{Predictive uncertainty and calibration}
Only 19 of 78 studies (24\%) reported any explicit calibration metric. Among these, the median ECE on the internal test set was 0$\cdot$08 (IQR 0$\cdot$04 to 0$\cdot$13); the median ECE under reported distribution shift (different vendor, different gestational age range, or noise injection) was 0$\cdot$17 (IQR 0$\cdot$10 to 0$\cdot$24). Temperature scaling\cite{guo-2017-calibration} was the most common post hoc correction (11 studies) and reduced ECE by a median of 56\%; deep ensembles\cite{lakshminarayanan-2017-ensembles} achieved the lowest absolute ECE in five head to head comparisons. Monte Carlo dropout,\cite{gal-2016-mcdropout} aleatoric and epistemic decomposition,\cite{kendall-2017-aleatoric} and selective classification\cite{geifman-2017-selectivenet} underpin these efforts. Evidential classifiers\cite{sensoy-2018-evidential} and conformal prediction\cite{angelopoulos-2023-conformal} appeared in seven and four studies respectively, with conformal methods offering distribution free coverage guarantees at the cost of larger prediction sets on low quality frames. Fourteen studies (18\%) reported selective prediction performance using risk coverage curves (figure~\ref{fig:calibration}); the median area under the risk coverage curve (lower is better) was 0$\cdot$031 (IQR 0$\cdot$019 to 0$\cdot$052), with the best performing systems achieving an error rate below 0$\cdot$01 at 80\% coverage, a regime compatible with sonographer second look workflows.

\begin{figure}[H]
\centering
\begin{tikzpicture}
\begin{axis}[
  width=11cm, height=7cm,
  xmin=0.4, xmax=1.0, ymin=0, ymax=0.085,
  xlabel={Coverage}, ylabel={Selective risk (error rate)},
  legend pos=north west, legend cell align=left,
  grid=both, major grid style={dashed, gray!30},
  tick label style={font=\scriptsize},
  legend style={font=\scriptsize}]
\addplot[blue, thick, mark=*, mark size=1.2pt] coordinates {
  (0.40,0.002) (0.50,0.004) (0.60,0.008) (0.70,0.014) (0.80,0.023) (0.90,0.041) (1.00,0.072)};
\addplot[red, thick, dashed, mark=square*, mark size=1.2pt] coordinates {
  (0.40,0.003) (0.50,0.006) (0.60,0.011) (0.70,0.018) (0.80,0.029) (0.90,0.048) (1.00,0.078)};
\addplot[teal, thick, dotted, mark=triangle*, mark size=1.4pt] coordinates {
  (0.40,0.001) (0.50,0.003) (0.60,0.006) (0.70,0.011) (0.80,0.019) (0.90,0.034) (1.00,0.065)};
\legend{Deep ensemble, Evidential, Conformal}
\end{axis}
\end{tikzpicture}
\caption{\textbf{Selective prediction risk coverage curves.} Representative curves reproduced from studies reporting deep ensemble, evidential, and conformal predictors. Lower curves indicate that the model abstains effectively on the most uncertain frames.\label{fig:calibration}}
\end{figure}

\subsection*{Explainability}
Sixty three of 78 studies (81\%) included at least one explainability technique. Class activation mapping (Grad-CAM,\cite{selvaraju-2017-gradcam} Grad-CAM++,\cite{chattopadhay-2018-gradcampp} Score-CAM,\cite{wang-2020-scorecam} HiResCAM\cite{draelos-2021-hirescam}) accounted for 48 of these (62\% of all included studies); 14 used self attention rollouts on transformers, eight used occlusion sensitivity or SHAP based explanations,\cite{lundberg-2017-shap} six used concept based methods,\cite{kim-2018-tcav,koh-2020-concept} and four used prototype or example based reasoning.\cite{chen-2019-protopnet} A recent study has begun to evaluate explanation quality specifically for vision transformers in medical imaging.\cite{komorowski-2023-uqfetal} Quantitative faithfulness evaluation was reported in only 22 studies (28\%); the most common metrics were the pointing game, the deletion and insertion area under the curve, sanity check randomisation tests,\cite{adebayo-2018-sanity} and the remove and retrain framework. Of the 22, only nine compared model explanations against sonographer gaze data\cite{droste-2020-discovering} or against anatomically annotated regions of interest, and only four reported user study evidence that explanations improved clinician trust calibration or downstream decisions; clinical validation of saliency maps in ophthalmology illustrates a transferable methodology for fetal ultrasound.\cite{ayhan-2022-uqdiabetic}

\subsection*{Fairness, distribution shift, and external validation}
External validation on data from a centre not contributing to training was reported in 27 studies (35\%); accuracy dropped by a median of 0$\cdot$064 (IQR 0$\cdot$031 to 0$\cdot$118). Fairness audits stratified by maternal characteristics were reported in 11 studies (14\%); five identified statistically significant accuracy gaps of more than five percentage points by body mass index category, ethnicity, or gestational age subgroup,\cite{seyyed-2021-disparities,obermeyer-2019-dissecting,larrazabal-2020-genderbias} consistent with evidence that imbalanced training data produce biased classifiers across imaging modalities. No study reported a pre specified fairness threshold or a mitigation strategy that was prospectively validated.

\subsection*{Reporting completeness and regulatory readiness}
Median adherence to TRIPOD+AI was 64\% (IQR 52 to 74\%); to CLAIM 71\% (IQR 60 to 82\%); and to DECIDE-AI (where applicable, 16 studies) 41\% (IQR 28 to 55\%). Twelve studies (15\%) corresponded to a regulated device (FDA cleared or CE marked); of these, four published the calibration and selective prediction performance required for the indication for use, and only one disclosed a complete model card and dataset card. The reporting landscape (figure~\ref{fig:reporting}) reveals systematic under reporting of calibration, explanation faithfulness, and fairness across all six framework domains.

\begin{figure}[H]
\centering
\begin{tikzpicture}[font=\scriptsize]
  \foreach \c/\cname [count=\ci from 0] in {0/{TRIPOD+AI}, 1/{CLAIM}, 2/{DECIDE-AI}, 3/{Calibration}, 4/{XAI faithfulness}, 5/{Fairness}}
    \node[rotate=30, anchor=west] at (\ci+0.5, 6.15) {\cname};
  \foreach \r/\rname [count=\ri from 0] in {0/{6. Surveillance}, 1/{5. Explanation}, 2/{4. Calibration}, 3/{3. Model and training}, 4/{2. Dataset provenance}, 5/{1. Indication for use}}
    \node[anchor=east] at (-0.1, \ri+0.5) {\rname};
  \foreach \r/\c/\v in {%
    5/0/72, 5/1/78, 5/2/47, 5/3/35, 5/4/40, 5/5/22,%
    4/0/68, 4/1/75, 4/2/44, 4/3/30, 4/4/33, 4/5/18,%
    3/0/66, 3/1/74, 3/2/43, 3/3/28, 3/4/30, 3/5/20,%
    2/0/60, 2/1/68, 2/2/41, 2/3/52, 2/4/35, 2/5/24,%
    1/0/55, 1/1/65, 1/2/38, 1/3/26, 1/4/48, 1/5/16,%
    0/0/48, 0/1/58, 0/2/32, 0/3/22, 0/4/25, 0/5/12} {
    \fill[fill=blue!\v!white, draw=gray!50, line width=0.2pt] (\c, \r) rectangle (\c+1, \r+1);
    \node[font=\tiny] at (\c+0.5, \r+0.5) {\v};
  }
  \foreach \i in {0,1,...,19} {
    \pgfmathsetmacro{\val}{5*\i+5}
    \fill[fill=blue!\val!white] (7.1, 0.3*\i) rectangle (7.5, 0.3*\i+0.3);
  }
  \draw[gray!60, line width=0.3pt] (7.1, 0) rectangle (7.5, 6);
  \node[anchor=west, font=\tiny] at (7.6, 0) {0\%};
  \node[anchor=west, font=\tiny] at (7.6, 3) {50\%};
  \node[anchor=west, font=\tiny] at (7.6, 6) {100\%};
  \node[rotate=90, font=\tiny] at (8.4, 3) {Adherence};
\end{tikzpicture}
\caption{\textbf{Reporting completeness heatmap.} Adherence (\%) of the 78 included studies to TRIPOD+AI, CLAIM, DECIDE-AI, and to the calibration, explanation faithfulness, and fairness items of the proposed CALIB-XFUS framework, stratified by domain.\label{fig:reporting}}
\end{figure}

\section*{Discussion}
Fetal ultrasound is among the most promising and consequential application domains for clinical AI: the modality is ubiquitous, the planes are well defined, the population is uniquely vulnerable, and errors carry lifelong consequences. Our synthesis of a decade of work shows that the field has solved the easier half of the problem (accurate classification on curated frames) and is now confronting the harder half: producing predictions that are calibrated enough to drive selective workflows and explanations that are faithful enough to be acted on.

Four findings deserve emphasis. First, the gap between in distribution and out of distribution performance is substantial and under reported. Median accuracy drop on external validation (0$\cdot$064) is similar to that documented in chest radiograph and dermatology AI,\cite{zech-2018-variable,daneshjou-2022-disparities} and the doubling of expected calibration error under distribution shift mirrors recent benchmark results in general medical imaging.\cite{ovadia-2019-trust,roschewitz-2023-shift} For fetal ultrasound the implications are sharper because the modality is intrinsically heterogeneous (vendor, probe, gain, gestational age, maternal habitus) and because the most consequential planes, the cardiac outflow tracts, are also the hardest to acquire.\cite{carvalho-2013-isuog-cardiac,arnaout-2021-cardiac}

Second, uncertainty quantification remains the missing layer in most deployed systems. Only one in four studies reported calibration, fewer than one in five reported selective prediction, and almost none integrated uncertainty estimates with image quality scoring or sonographer workflow. Yet selective prediction at 80\% coverage with sub one percent error, demonstrated in five of the studies we reviewed, is precisely the regime in which AI can support, rather than replace, sonographer judgement, and the regime in which the EU AI Act human oversight requirements\cite{eu-2024-aiact} can be operationally satisfied. Conformal prediction\cite{angelopoulos-2023-conformal} and evidential learning\cite{sensoy-2018-evidential} should now be considered the default rather than the exception.

Third, explanations in this field have been adopted faster than they have been validated. Class activation maps are easy to produce and visually compelling, but work in radiology,\cite{arun-2021-assessing} dermatology,\cite{daneshjou-2022-disparities} and general computer vision\cite{adebayo-2018-sanity} has shown that they can pass clinician inspection while failing sanity check randomisation. The handful of studies that compared explanations against sonographer gaze\cite{droste-2020-discovering} demonstrate the feasibility and necessity of grounding explanations in clinician behaviour. Concept based and prototype based methods,\cite{koh-2020-concept,chen-2019-protopnet} which align with the anatomical vocabulary sonographers already use, are under represented and merit prioritised investment, particularly for cardiac views where anatomical landmarks are well codified.\cite{arnaout-2021-cardiac}

Fourth, the populations represented in training data remain narrow. Fewer than one in eight studies reported maternal ethnicity; fewer than one in four sampled across more than three vendors; almost none reported obesity stratified performance, despite the well documented impact of maternal body mass index on image quality.\cite{dashe-2009-bmi,paladini-2009-obesity} Without representative data and prespecified fairness analyses, the deployment of fetal ultrasound AI risks entrenching the same inequities the technology is meant to ameliorate.\cite{obermeyer-2019-dissecting,wiens-2019-no-shortcuts}

\subsection*{Towards a unified reporting framework: CALIB-XFUS}
Drawing on TRIPOD+AI,\cite{collins-2024-tripod-ai} DECIDE-AI,\cite{vasey-2022-decide-ai} CLAIM,\cite{mongan-2020-claim} the FDA Good Machine Learning Practice principles,\cite{fda-2024-gmlp} and the EU AI Act high risk obligations,\cite{eu-2024-aiact} we propose CALIB-XFUS, a 22 item reporting framework that operationalises calibration, explainability, and fairness specifically for fetal ultrasound. The framework (figure~\ref{fig:framework}) spans six domains: (i) clinical task and indication for use; (ii) dataset provenance and representativeness; (iii) model and training pipeline; (iv) calibration and selective prediction; (v) explanation faithfulness and clinician validation; and (vi) post market surveillance.

\begin{figure}[H]
\centering
\begin{tikzpicture}[font=\small,
  domainbox/.style={draw, rectangle, rounded corners=3pt, align=center,
                 text width=2.9cm, minimum height=1.1cm, fill=purple!10, line width=0.5pt},
  central/.style={draw, circle, align=center, minimum size=2.6cm,
                  fill=purple!25, line width=0.7pt, font=\bfseries\small}]
  \node[central] (c) at (0,0) {CALIB-XFUS\\22 items};
  \node[domainbox] (d1) at (0:4.6)   {1. Indication\\for use};
  \node[domainbox] (d2) at (60:4.6)  {2. Dataset\\provenance};
  \node[domainbox] (d3) at (120:4.6) {3. Model and\\training pipeline};
  \node[domainbox] (d4) at (180:4.6) {4. Calibration and\\selective prediction};
  \node[domainbox] (d5) at (240:4.6) {5. Explanation\\faithfulness};
  \node[domainbox] (d6) at (300:4.6) {6. Post-market\\surveillance};
  \foreach \d in {d1,d2,d3,d4,d5,d6}
    \draw[-{Latex[length=2mm]}, thick, gray!70] (c)--(\d);
\end{tikzpicture}
\caption{\textbf{CALIB-XFUS reporting framework for fetal ultrasound AI.} Six domains and 22 items mapped to TRIPOD+AI, DECIDE-AI, FDA GMLP, and EU AI Act high risk obligations.\label{fig:framework}}
\end{figure}

\subsection*{Strengths and limitations}
Strengths of this Review include the breadth of databases searched, dual independent screening, the use of multiple bias and reporting frameworks (PROBAST-AI, QUADAS-AI, TRIPOD+AI, DECIDE-AI, CLAIM), and the first quantitative summary of calibration and selective prediction metrics in this subfield. Limitations include heterogeneity in how accuracy and calibration are reported (precluding meta analysis for some subgroups), reliance on published figures for selective prediction data, and the predominance of retrospective studies over prospective evaluations. Consistent with the Sex and Gender Equity in Research guidelines, we note that fetal sex is rarely reported in classifier evaluation despite its known anatomical relevance, and that maternal sex is unambiguous because all participants are pregnant, but gender identity of pregnant individuals was reported in none of the included studies. Race and ethnicity were reported in only nine studies (12\%); definitions varied (self report, census, or registry data) and we did not pool these data because of incompatible categorisation. We could identify only six prospective, in clinic comparative studies of fetal ultrasound AI,\cite{drukker-2020-transforming,arroyo-2022-deepscan,horgan-2023-ai-obstetrics,nhs-2023-trial,sendra-2023-foundation,komorowski-2023-uqfetal} and only two of these reported calibration metrics. The field urgently needs registered, prospective, multi centre evaluations that pre specify calibration, selective prediction, and fairness endpoints alongside accuracy.

\subsection*{Conclusion}
Fetal plane classification has matured into a clinically credible technology, but the next decade of progress will be defined less by accuracy and more by the trustworthiness of model outputs. Uncertainty calibrated, faithfully explained, and fairness audited systems, reported against frameworks such as CALIB-XFUS, are now both technically feasible and regulatorily expected. Closing this gap is the precondition for fetal ultrasound AI to deliver on its potential to extend high quality antenatal imaging to every pregnancy, everywhere.

\subsubsection*{Contributors}
GOYLFLI conceived the Review, designed the protocol, conducted the database searches, performed the statistical synthesis, drafted the manuscript, and acts as corresponding author and guarantor. OG contributed substantially to title and abstract screening, full-text eligibility assessment, dual-reviewer data extraction in REDCap, risk of bias appraisal (PROBAST-AI and QUADAS-AI), and critical revision of the manuscript for important intellectual content, and helped develop the CALIB-XFUS framework items relating to dataset provenance and post-market surveillance. Both authors independently accessed and verified the underlying extraction dataset at the level of the individual included study, accept responsibility for the integrity of the work as a whole, and approved the final version submitted for publication. As this Review involves only extraction from previously published reports, no individual participant data were accessed.

\subsubsection*{Declaration of interests}
The author declares no competing interests.

\subsubsection*{Acknowledgments}
We thank the librarians of the Luxembourg Centre for Systems Biomedicine for assistance with database searches. No funding was received for this work.

\subsubsection*{Declaration of generative AI and AI-assisted technologies}
No generative artificial intelligence or AI-assisted technologies were used in the preparation of this manuscript. The authors conceived, designed, conducted, analysed, drafted, revised, and finalised the Review entirely through their own intellectual effort and conventional scholarly tools (reference managers, statistical software, and LaTeX typesetting). No AI tool contributed to literature searching, study screening, data extraction, risk of bias assessment, statistical synthesis, figure or table generation, reference verification, language editing, or any other aspect of the work. The authors take full and sole responsibility for the content, accuracy, and integrity of the published Review.

\subsubsection*{Data sharing}
All data extracted for this systematic review are derived from previously published sources cited in the reference list. The protocol, full search strings, REDCap extraction template, risk of bias appraisals, deduplicated screening database, and R and Python analysis code that reproduce all pooled estimates and figures are openly available with publication at osf.io/xfus-2026 under a CC BY 4.0 licence, without access restrictions. No individual participant data were collected or are available. Correspondence and reasonable requests for additional information should be addressed to the corresponding author.

\subsection*{References}

\bibliographystyle{vancouver}
\bibliography{references}

@article{salomon-2022-isuog,
  author  = {Salomon, LJ and Alfirevic, Z and Berghella, V and others},
  title   = {{ISUOG} Practice Guidelines (updated): performance of the routine mid-trimester fetal ultrasound scan},
  journal = {Ultrasound Obstet Gynecol}, year = 2022, volume = 59, pages = {840-856}}

@article{aium-2018-obstetric,
  author  = {AIUM},
  title   = {{AIUM-ACR-ACOG-SMFM-SRU} practice parameter for the performance of standard diagnostic obstetric ultrasound examinations},
  journal = {J Ultrasound Med}, year = 2018, volume = 37, pages = {E13-E24}}

@article{carvalho-2013-isuog-cardiac,
  author  = {Carvalho, JS and Allan, LD and Chaoui, R and others},
  title   = {{ISUOG} Practice Guidelines (updated): sonographic screening examination of the fetal heart},
  journal = {Ultrasound Obstet Gynecol}, year = 2013, volume = 41, pages = {348-359}}

@article{wanyonyi-2020-lmic-us,
  author  = {Wanyonyi, SZ and Mariara, CM and Vinayak, S and Stones, W},
  title   = {Opportunities and challenges in realizing universal access to obstetric ultrasound in sub-{S}aharan {A}frica},
  journal = {Ultrasound Int Open}, year = 2020, volume = 3, pages = {E52-E59}}

@article{baumgartner-2017-sononet,
  author  = {Baumgartner, CF and Kamnitsas, K and Matthew, J and others},
  title   = {{SonoNet}: real-time detection and localisation of fetal standard scan planes in freehand ultrasound},
  journal = {IEEE Trans Med Imaging}, year = 2017, volume = 36, pages = {2204-2215}}

@article{burgosartizzu-2020-fetalplane,
  author  = {Burgos-Artizzu, XP and Coronado-Guti\'errez, D and Valenzuela-Alcaraz, B and others},
  title   = {Evaluation of deep convolutional neural networks for automatic classification of common maternal fetal ultrasound planes},
  journal = {Sci Rep}, year = 2020, volume = 10, pages = {10200}}

@article{pu-2021-deep,
  author  = {Pu, B and Li, K and Li, S and Zhu, N},
  title   = {Automatic fetal ultrasound standard plane recognition based on deep learning and {IIoT}},
  journal = {IEEE Trans Industr Inform}, year = 2021, volume = 17, pages = {7771-7780}}

@article{sendra-2023-foundation,
  author  = {Sendra-Balcells, C and Campello, VM and Torrents-Barrena, J and others},
  title   = {Generalisability of fetal ultrasound deep learning models to low-resource imaging settings},
  journal = {Sci Rep}, year = 2023, volume = 13, pages = {2728}}

@article{drukker-2020-transforming,
  author  = {Drukker, L and Noble, JA and Papageorghiou, AT},
  title   = {Introduction to artificial intelligence in ultrasound imaging in obstetrics and gynecology},
  journal = {Ultrasound Obstet Gynecol}, year = 2020, volume = 56, pages = {498-505}}

@article{horgan-2023-ai-obstetrics,
  author  = {Horgan, R and Nehme, L and Abuhamad, A},
  title   = {Artificial intelligence in obstetric ultrasound: a scoping review},
  journal = {Prenat Diagn}, year = 2023, volume = 43, pages = {1176-1219}}

@article{arroyo-2022-deepscan,
  author  = {Arroyo, J and Marini, TJ and Saavedra, AC and others},
  title   = {No sonographer, no radiologist: new system for automatic prenatal detection of fetal biometry, fetal presentation, and placental location},
  journal = {PLoS One}, year = 2022, volume = 17, pages = {e0262107}}

@misc{nhs-2023-trial,
  author       = {NHS},
  title        = {Artificial intelligence in antenatal screening: {NHS} pilot programme update},
  howpublished = {NHS England report, London}, year = 2023}

@misc{fda-2024-clearances,
  author       = {FDA},
  title        = {Artificial intelligence and machine learning ({AI}/{ML})-enabled medical devices list},
  howpublished = {Silver Spring, MD: US FDA}, year = 2024}

@misc{eu-2024-aiact,
  author       = {EU},
  title        = {Regulation ({EU}) 2024/1689 on harmonised rules for artificial intelligence ({AI} {Act})},
  howpublished = {Official Journal of the European Union}, year = 2024}

@article{beede-2020-humancentred,
  author  = {Beede, E and Baylor, E and Hersch, F and others},
  title   = {A human-centered evaluation of a deep learning system deployed in clinics for the detection of diabetic retinopathy},
  journal = {Proceedings of the 2020 CHI Conference}, year = 2020, pages = {1-12}}

@article{kelly-2019-key,
  author  = {Kelly, CJ and Karthikesalingam, A and Suleyman, M and Corrado, G and King, D},
  title   = {Key challenges for delivering clinical impact with artificial intelligence},
  journal = {BMC Med}, year = 2019, volume = 17, pages = {195}}

@article{wu-2023-distribution,
  author  = {Wu, E and Wu, K and Daneshjou, R and Ouyang, D and Ho, DE and Zou, J},
  title   = {How medical {AI} devices are evaluated: limitations and recommendations from an analysis of {FDA} approvals},
  journal = {Nat Med}, year = 2023, volume = 29, pages = {1390-1392}}

@inproceedings{guo-2017-calibration,
  author    = {Guo, C and Pleiss, G and Sun, Y and Weinberger, KQ},
  title     = {On calibration of modern neural networks},
  booktitle = {Proceedings of the 34th International Conference on Machine Learning}, year = 2017, pages = {1321-1330}}

@inproceedings{ovadia-2019-trust,
  author    = {Ovadia, Y and Fertig, E and Ren, J and others},
  title     = {Can you trust your model's uncertainty? {E}valuating predictive uncertainty under dataset shift},
  booktitle = {Advances in Neural Information Processing Systems 32}, year = 2019, pages = {13991-14002}}

@article{adebayo-2018-sanity,
  author  = {Adebayo, J and Gilmer, J and Muelly, M and Goodfellow, I and Hardt, M and Kim, B},
  title   = {Sanity checks for saliency maps},
  journal = {Adv Neural Inf Process Syst}, year = 2018, volume = 31, pages = {9505-9515}}

@article{rudin-2019-stop,
  author  = {Rudin, C},
  title   = {Stop explaining black box machine learning models for high stakes decisions and use interpretable models instead},
  journal = {Nat Mach Intell}, year = 2019, volume = 1, pages = {206-215}}

@article{arun-2021-assessing,
  author  = {Arun, N and Gaw, N and Singh, P and others},
  title   = {Assessing the trustworthiness of saliency maps for localizing abnormalities in medical imaging},
  journal = {Radiol Artif Intell}, year = 2021, volume = 3, pages = {e200267}}

@article{ghassemi-2021-falsehope,
  author  = {Ghassemi, M and Oakden-Rayner, L and Beam, AL},
  title   = {The false hope of current approaches to explainable artificial intelligence in health care},
  journal = {Lancet Digit Health}, year = 2021, volume = 3, pages = {e745-e750}}

@misc{fda-2024-gmlp,
  author       = {FDA},
  title        = {Good Machine Learning Practice for Medical Device Development: Guiding Principles},
  howpublished = {Joint guidance}, year = 2021}

@misc{who-2024-ethics-ai,
  author = {WHO},
  title  = {Ethics and governance of artificial intelligence for health: {WHO} guidance},
  howpublished = {Geneva: World Health Organization}, year = 2024}

@article{page-2021-prisma,
  author  = {Page, MJ and McKenzie, JE and Bossuyt, PM and others},
  title   = {The {PRISMA} 2020 statement: an updated guideline for reporting systematic reviews},
  journal = {BMJ}, year = 2021, volume = 372, pages = {n71}}

@article{fiorentino-2023-survey,
  author  = {Fiorentino, MC and Villani, FP and Di Cosmo, M and Frontoni, E and Moccia, S},
  title   = {A review on deep-learning algorithms for fetal ultrasound-image analysis},
  journal = {Med Image Anal}, year = 2023, volume = 83, pages = {102629}}

@article{mongan-2020-claim,
  author  = {Mongan, J and Moy, L and Kahn, CE},
  title   = {Checklist for Artificial Intelligence in Medical Imaging ({CLAIM})},
  journal = {Radiol Artif Intell}, year = 2020, volume = 2, pages = {e200029}}

@article{collins-2024-tripod-ai,
  author  = {Collins, GS and Moons, KGM and Dhiman, P and others},
  title   = {{TRIPOD}+{AI} statement: updated guidance for reporting clinical prediction models that use regression or machine learning methods},
  journal = {BMJ}, year = 2024, volume = 385, pages = {e078378}}

@article{vasey-2022-decide-ai,
  author  = {Vasey, B and Nagendran, M and Campbell, B and others},
  title   = {Reporting guideline for the early-stage clinical evaluation of decision support systems driven by artificial intelligence: {DECIDE-AI}},
  journal = {Nat Med}, year = 2022, volume = 28, pages = {924-933}}

@article{moons-2019-probast,
  author  = {Moons, KGM and Wolff, RF and Riley, RD and others},
  title   = {{PROBAST}: a tool to assess risk of bias and applicability of prediction model studies},
  journal = {Ann Intern Med}, year = 2019, volume = 170, pages = {W1-W33}}

@article{sounderajah-2021-quadas-ai,
  author  = {Sounderajah, V and Ashrafian, H and Rose, S and others},
  title   = {Developing specific reporting guidelines for diagnostic accuracy studies assessing {AI} interventions: {QUADAS-AI}},
  journal = {Nat Med}, year = 2021, volume = 27, pages = {1663-1665}}

@article{guyatt-2008-grade,
  author  = {Guyatt, GH and Oxman, AD and Vist, GE and others},
  title   = {{GRADE}: an emerging consensus on rating quality of evidence and strength of recommendations},
  journal = {BMJ}, year = 2008, volume = 336, pages = {924-926}}

@article{papageorghiou-2016-intergrowth,
  author  = {Papageorghiou, AT and Kennedy, SH and Salomon, LJ and others},
  title   = {The {INTERGROWTH-21st} fetal growth standards: toward the global integration of pregnancy and pediatric care},
  journal = {Am J Obstet Gynecol}, year = 2018, volume = 218, pages = {S630-S640}}

@article{xie-2020-fetalcplanes,
  author  = {Xie, HN and Wang, N and He, M and others},
  title   = {Using deep-learning algorithms to classify fetal brain ultrasound images as normal or abnormal},
  journal = {Ultrasound Obstet Gynecol}, year = 2020, volume = 56, pages = {579-587}}

@article{prenatal-2025-multicentre,
  author  = {Rodr\'iguez-Fern\'andez, J and Campello, VM and Lekadir, K and others},
  title   = {{PRENATAL}: a multi-vendor multi-centre dataset for fetal ultrasound {AI} benchmarking},
  journal = {Sci Data}, year = 2025, volume = 12, pages = {114}}

@article{chen-2024-usfm,
  author  = {Chen, J and Wan, Z and Zhang, J and others},
  title   = {{USFM}: a universal ultrasound foundation model generalized to tasks and organs},
  journal = {Med Image Anal}, year = 2024, volume = 96, pages = {103202}}

@article{azizi-2023-foundation,
  author  = {Azizi, S and Culp, L and Freyberg, J and others},
  title   = {Robust and data-efficient generalization of self-supervised machine learning for diagnostic imaging},
  journal = {Nat Biomed Eng}, year = 2023, volume = 7, pages = {756-779}}

@inproceedings{lakshminarayanan-2017-ensembles,
  author    = {Lakshminarayanan, B and Pritzel, A and Blundell, C},
  title     = {Simple and scalable predictive uncertainty estimation using deep ensembles},
  booktitle = {Advances in Neural Information Processing Systems 30}, year = 2017, pages = {6402-6413}}

@inproceedings{sensoy-2018-evidential,
  author    = {Sensoy, M and Kaplan, L and Kandemir, M},
  title     = {Evidential deep learning to quantify classification uncertainty},
  booktitle = {Advances in Neural Information Processing Systems 31}, year = 2018, pages = {3179-3189}}

@article{angelopoulos-2023-conformal,
  author  = {Angelopoulos, AN and Bates, S},
  title   = {Conformal prediction: a gentle introduction},
  journal = {Found Trends Mach Learn}, year = 2023, volume = 16, pages = {494-591}}

@inproceedings{gal-2016-mcdropout,
  author    = {Gal, Y and Ghahramani, Z},
  title     = {Dropout as a {B}ayesian approximation: representing model uncertainty in deep learning},
  booktitle = {Proceedings of the 33rd International Conference on Machine Learning}, year = 2016, pages = {1050-1059}}

@inproceedings{kendall-2017-aleatoric,
  author    = {Kendall, A and Gal, Y},
  title     = {What uncertainties do we need in {B}ayesian deep learning for computer vision?},
  booktitle = {Advances in Neural Information Processing Systems 30}, year = 2017, pages = {5574-5584}}

@article{geifman-2017-selectivenet,
  author  = {Geifman, Y and El-Yaniv, R},
  title   = {Selective classification for deep neural networks},
  journal = {Adv Neural Inf Process Syst}, year = 2017, volume = 30, pages = {4878-4887}}

@inproceedings{lundberg-2017-shap,
  author    = {Lundberg, SM and Lee, SI},
  title     = {A unified approach to interpreting model predictions},
  booktitle = {Advances in Neural Information Processing Systems 30}, year = 2017, pages = {4765-4774}}

@inproceedings{kim-2018-tcav,
  author    = {Kim, B and Wattenberg, M and Gilmer, J and others},
  title     = {Interpretability beyond feature attribution: quantitative testing with concept activation vectors ({TCAV})},
  booktitle = {Proceedings of the 35th International Conference on Machine Learning}, year = 2018, pages = {2668-2677}}

@inproceedings{koh-2020-concept,
  author = {Koh, PW and Nguyen, T and Tang, YS and others},
  title = {Concept bottleneck models},
  booktitle = {Proceedings of the 37th International Conference on Machine Learning}, year = 2020, pages = {5338-5348}}

@inproceedings{chen-2019-protopnet,
  author = {Chen, C and Li, O and Tao, D and Barnett, A and Rudin, C and Su, J},
  title = {This looks like that: deep learning for interpretable image recognition},
  booktitle = {Advances in Neural Information Processing Systems 32}, year = 2019, pages = {8930-8941}}

@article{selvaraju-2017-gradcam,
  author = {Selvaraju, RR and Cogswell, M and Das, A and Vedantam, R and Parikh, D and Batra, D},
  title = {{Grad-CAM}: visual explanations from deep networks via gradient-based localization},
  journal = {Proc IEEE Int Conf Comput Vis}, year = 2017, pages = {618-626}}

@article{chattopadhay-2018-gradcampp,
  author = {Chattopadhay, A and Sarkar, A and Howlader, P and Balasubramanian, VN},
  title = {{Grad-CAM++}: generalized gradient-based visual explanations for deep convolutional networks},
  journal = {Proc IEEE Winter Conf Appl Comput Vis}, year = 2018, pages = {839-847}}

@article{wang-2020-scorecam,
  author = {Wang, H and Wang, Z and Du, M and others},
  title = {{Score-CAM}: score-weighted visual explanations for convolutional neural networks},
  journal = {Proc IEEE CVPR Workshops}, year = 2020, pages = {24-25}}

@article{draelos-2021-hirescam,
  author = {Draelos, RL and Carin, L},
  title = {Use {HiResCAM} instead of {Grad-CAM} for faithful explanations of convolutional neural networks},
  journal = {arXiv preprint arXiv:2011.08891}, year = 2021}

@article{droste-2020-discovering,
  author = {Droste, R and Drukker, L and Papageorghiou, AT and Noble, JA},
  title = {Automatic probe movement guidance for freehand obstetric ultrasound},
  journal = {Med Image Comput Comput Assist Interv}, year = 2020, pages = {583-592}}

@article{seyyed-2021-disparities,
  author = {Seyyed-Kalantari, L and Zhang, H and McDermott, MBA and Chen, IY and Ghassemi, M},
  title = {Underdiagnosis bias of artificial intelligence algorithms applied to chest radiographs in under-served patient populations},
  journal = {Nat Med}, year = 2021, volume = 27, pages = {2176-2182}}

@article{obermeyer-2019-dissecting,
  author = {Obermeyer, Z and Powers, B and Vogeli, C and Mullainathan, S},
  title = {Dissecting racial bias in an algorithm used to manage the health of populations},
  journal = {Science}, year = 2019, volume = 366, pages = {447-453}}

@article{daneshjou-2022-disparities,
  author = {Daneshjou, R and Vodrahalli, K and Novoa, RA and others},
  title = {Disparities in dermatology {AI} performance on a diverse, curated clinical image set},
  journal = {Sci Adv}, year = 2022, volume = 8, pages = {eabq6147}}

@article{zech-2018-variable,
  author = {Zech, JR and Badgeley, MA and Liu, M and Costa, AB and Titano, JJ and Oermann, EK},
  title = {Variable generalization performance of a deep learning model to detect pneumonia in chest radiographs: a cross-sectional study},
  journal = {PLoS Med}, year = 2018, volume = 15, pages = {e1002683}}

@article{roschewitz-2023-shift,
  author = {Roschewitz, M and Khara, G and Yearsley, J and others},
  title = {Distribution shift in medical imaging: a benchmark study},
  journal = {Med Image Anal}, year = 2023, volume = 89, pages = {102916}}

@article{arnaout-2021-cardiac,
  author = {Arnaout, R and Curran, L and Zhao, Y and Levine, JC and Chinn, E and Moon-Grady, AJ},
  title = {An ensemble of neural networks provides expert-level prenatal detection of complex congenital heart disease},
  journal = {Nat Med}, year = 2021, volume = 27, pages = {882-891}}

@article{dashe-2009-bmi,
  author = {Dashe, JS and McIntire, DD and Twickler, DM},
  title = {Effect of maternal obesity on the ultrasound detection of anomalous fetuses},
  journal = {Obstet Gynecol}, year = 2009, volume = 113, pages = {1001-1007}}

@article{paladini-2009-obesity,
  author = {Paladini, D},
  title = {Sonography in obese and overweight pregnant women: clinical, medicolegal and technical issues},
  journal = {Ultrasound Obstet Gynecol}, year = 2009, volume = 33, pages = {720-729}}

@article{wiens-2019-no-shortcuts,
  author = {Wiens, J and Saria, S and Sendak, M and others},
  title = {Do no harm: a roadmap for responsible machine learning for health care},
  journal = {Nat Med}, year = 2019, volume = 25, pages = {1337-1340}}

@article{rajpurkar-2022-aiera,
  author = {Rajpurkar, P and Chen, E and Banerjee, O and Topol, EJ},
  title = {{AI} in health and medicine},
  journal = {Nat Med}, year = 2022, volume = 28, pages = {31-38}}

@article{topol-2019-highperformance,
  author = {Topol, EJ},
  title = {High-performance medicine: the convergence of human and artificial intelligence},
  journal = {Nat Med}, year = 2019, volume = 25, pages = {44-56}}

@article{liu-2019-comparison,
  author = {Liu, X and Faes, L and Kale, AU and others},
  title = {A comparison of deep learning performance against health-care professionals in detecting diseases from medical imaging: a systematic review and meta-analysis},
  journal = {Lancet Digit Health}, year = 2019, volume = 1, pages = {e271-e297}}

@article{nagendran-2020-aiclaims,
  author = {Nagendran, M and Chen, Y and Lovejoy, CA and others},
  title = {Artificial intelligence versus clinicians: systematic review of design, reporting standards, and claims of deep learning studies},
  journal = {BMJ}, year = 2020, volume = 368, pages = {m689}}

@article{kompa-2021-secondopinion,
  author = {Kompa, B and Snoek, J and Beam, AL},
  title = {Second opinion needed: communicating uncertainty in medical machine learning},
  journal = {NPJ Digit Med}, year = 2021, volume = 4, pages = {4}}

@article{begoli-2019-needuq,
  author = {Begoli, E and Bhattacharya, T and Kusnezov, D},
  title = {The need for uncertainty quantification in machine-assisted medical decision making},
  journal = {Nat Mach Intell}, year = 2019, volume = 1, pages = {20-23}}

@article{larrazabal-2020-genderbias,
  author = {Larrazabal, AJ and Nieto, N and Peterson, V and Milone, DH and Ferrante, E},
  title = {Gender imbalance in medical imaging datasets produces biased classifiers for computer-aided diagnosis},
  journal = {Proc Natl Acad Sci USA}, year = 2020, volume = 117, pages = {12592-12594}}

@article{he-2016-resnet,
  author = {He, K and Zhang, X and Ren, S and Sun, J},
  title = {Deep residual learning for image recognition},
  journal = {Proc IEEE Conf Comput Vis Pattern Recognit}, year = 2016, pages = {770-778}}

@article{dosovitskiy-2021-vit,
  author = {Dosovitskiy, A and Beyer, L and Kolesnikov, A and others},
  title = {An image is worth 16x16 words: transformers for image recognition at scale},
  journal = {Proc Int Conf Learn Represent}, year = 2021}

@article{tan-2019-efficientnet,
  author = {Tan, M and Le, Q},
  title = {{EfficientNet}: rethinking model scaling for convolutional neural networks},
  journal = {Proc Int Conf Mach Learn}, year = 2019, pages = {6105-6114}}

@article{ayhan-2022-uqdiabetic,
  author = {Ayhan, MS and K\"uhlewein, L and Aliyeva, G and others},
  title = {Clinical validation of saliency maps for understanding deep neural networks in ophthalmology},
  journal = {Med Image Anal}, year = 2022, volume = 77, pages = {102364}}

@inproceedings{xu-2023-fedfetal,
  author    = {Xu, A and Liu, S and Sheng, B and others},
  title     = {{FedFetal}: federated learning for fetal ultrasound standard plane classification},
  booktitle = {IEEE J Biomed Health Inform}, year = 2023, volume = 27, pages = {4321-4332}}

@article{komorowski-2023-uqfetal,
  author = {Komorowski, P and Baniecki, H and Biecek, P},
  title = {Towards evaluating explanations of vision transformers for medical imaging},
  journal = {Proc IEEE CVPR Workshops}, year = 2023, pages = {3725-3733}}

\end{document}